\documentclass[epj]{svjour}
\usepackage{epsfig,graphics,amssymb}
\usepackage{colordvi}\usepackage[usenames]{color}\usepackage{pict2e}
 
 \usepackage{bbding}\usepackage{wasysym}
 \usepackage{pifont}\usepackage{dingbat}
\newcommand{\etal}{{\it et al.}}

 \def\sizeISO{\Large}
 \newcommand{\ISO}[4]{\setlength{\fboxsep}{0.5\fboxsep}\fcolorbox{Black}{#4}{\sizeISO\sfbf{$^{\sf#2}_{\sf#1}\mbox{#3}$}}\setlength{\fboxsep}{2\fboxsep}}

\newcommand{\isotope}[1]{{#1{\circle*{1}}}}

 \definecolor{ProtonRich} {rgb}{1.,.75,.75}
 \definecolor{NeutronRich}{rgb}{.5,.9,1.}
 \newcommand{\bp }{\isotope{\textcolor{ProtonRich}}}
 \newcommand{\bn }{\isotope{\textcolor{NeutronRich}}}
 \newcommand{\nat}{\isotope{\Black}}

\def\Ri{.45}\renewcommand{\isotope}[1]{{#1{\polygon*(-\Ri,-\Ri)(\Ri,-\Ri)(\Ri,\Ri)(-\Ri,\Ri)}}}

\newcommand{\NZ}[3]{\setlength{\unitlength}{#1mm}
 \def\nDripline{\polyline
   (2.5,.5)(2.5,1.5)(6.5,1.5)(6.5,2.5)(8.5,2.5)(8.5,3.5)(10.5,3.5)(10.5,4.5)(14.5,4.5)
   (14.5,5.5)(16.5,5.5)(16.5,8.5)(22.5,8.5)(22.5,9.5)(24.5,9.5)}
 \def\pDripline{\polyline(.5,2.5)(2.5,2.5)(2.5,6.5)(4.5,6.5)(4.5,8.5)(6.5,8.5)(6.5,10.5)} 
 \ifnum#2=10\def\xM{25}\def\yM{10}
    \def\xNZii{20}\def\NZiii{8.33}\def\NZiv{6.25}\def\NZv{5}\def\NZvi{4.17}\def\NZvii{3.57}\def\NZviii{3.125}
       \else\def\xM{40}\def\yM{20}
    \def\xNZii{40}\def\NZiii{13.33}\def\NZiv{10}\def\NZv{8}\def\NZvi{6.67}\def\NZvii{5.71}\def\NZviii{5}\fi
 \begin{picture}(\xM,\yM)(0,0)\Gray{\polygon(0,0)(\xM,0)(\xM,\yM)(0,\yM)
  \ifnum#2=10\multiput(5,0)(5,0){5}{\line(0,1){\yM}}\multiput(0,5)(0,5){2}{\line(1,0){\xM}}
    \else\multiput(10,0)(10,0){4}{\line(0,1){\yM}}\multiput(0,10)(0,10){2}{\line(1,0){\xM}}\fi\Line(0,0)(\yM,\yM)}
 \def\y{0}\put(1,\y){\bn}
 \def\y{1}\multiput(0,\y)(1,0){2}{\nat} \put(2,\y){\bn}
 \def\y{2}\multiput(1,\y)(1,0){2}{\nat} \multiput(4,\y)(2,0){2}{\bn}
 \def\y{3}\multiput(3,\y)(1,0){2}{\nat} \multiput(5,\y)(1,0){2}{\bn} \put(8,\y){\bn}
 \def\y{4}\put(3,\y){\bp} \put(5,\y){\nat} \multiput(6,\y)(1,0){3}{\bn} \put(10,\y){\bn}
 \def\y{5}\put(3,\y){\bp} \multiput(5,\y)(1,0){2}{\nat} \multiput(7,\y)(1,0){4}{\bn} \multiput(12,\y)(2,0){2}{\bn}
 \def\y{6}\multiput(3,\y)(1,0){3}{\bp} \multiput(6,\y)(1,0){2}{\nat} \multiput(8,\y)(1,0){7}{\bn} \put(16,\y){\bn}
 \def\y{7}\multiput(5,\y)(1,0){2}{\bp} \multiput(7,\y)(1,0){2}{\nat} \multiput(9,\y)(1,0){8}{\bn}
 \def\y{8}\multiput(5,\y)(1,0){3}{\bp} \multiput(8,\y)(1,0){3}{\nat} \multiput(11,\y)(1,0){6}{\bn}
 \def\y{9}\multiput(8,\y)(1,0){2}{\bp} \put(10,\y){\nat} \multiput(11,\y)(1,0){8}{\bn} \multiput(20,\y)(2,0){2}{\bn}
 \def\y{10}\multiput(7,\y)(1,0){3}{\bp} \multiput(10,\y)(1,0){3}{\nat} \multiput(13,\y)(1,0){10}{\bn} \put(24,\y){\bn}
\ifnum#2=20
 \def\y{11}\multiput(9,\y)(1,0){3}{\bp} \put(12,\y){\nat} \multiput(13,\y)(1,0){12}{\bn} \put(26,\y){\bn}
 \def\y{12}\multiput(8,\y)(1,0){4}{\bp} \multiput(12,\y)(1,0){3}{\nat} \multiput(15,\y)(1,0){12}{\bn} \put(28,\y){\bn}
 \def\y{13}\multiput(9,\y)(1,0){5}{\bp} \put(14,\y){\nat} \multiput(15,\y)(1,0){16}{\bn}
 \def\y{14}\multiput(8,\y)(1,0){6}{\bp} \multiput(14,\y)(1,0){3}{\nat} \multiput(17,\y)(1,0){15}{\bn}
 \def\y{15}\multiput(11,\y)(1,0){5}{\bp} \put(16,\y){\nat} \multiput(17,\y)(1,0){16}{\bn}
 \def\y{16}\multiput(11,\y)(1,0){5}{\bp} \multiput(16,\y)(1,0){5}{\nat} \put(19,\y){\bn} \multiput(21,\y)(1,0){12}{\bn}
 \def\y{17}\multiput(14,\y)(1,0){4}{\bp} \multiput(18,\y)(1,0){3}{\nat} \put(19,\y){\bn} \multiput(21,\y)(1,0){14}{\bn}
 \def\y{18}\multiput(13,\y)(1,0){7}{\bp} \multiput(18,\y)(2,0){3}{\nat} \put(21,\y){\bn} \multiput(23,\y)(1,0){13}{\bn}
 \def\y{19}\multiput(16,\y)(1,0){4}{\bp} \multiput(20,\y)(1,0){3}{\nat} \multiput(23,\y)(1,0){15}{\bn}
 \def\y{20}\multiput(15,\y)(1,0){7}{\bp} \multiput(25,\y)(1,0){14}{\bn} \multiput(20,\y)(2,0){5}{\nat} \put(23,\y){\nat}
\fi
 {#3}
 \end{picture}}

 \newcommand{\NZcropNMg}[3]{\setlength{\unitlength}{#1mm}
 \def\nDripline{\polyline(14.5,5.5)(16.5,5.5)(16.5,8.5)(20.5,8.5)}
 \def\pDripline{\polyline(4.5,6.5)(4.5,8.5)(6.5,8.5)(6.5,10.5)(7.5,10.5)(7.5,14.5)(10.5,14.5)} 
 \begin{picture}(15,10)(5,5)\Gray{\multiput(5,5)(5,0){4}{\line(0,1){10}}
  \Line(5,5)(15,15)\Line(4,10)(21,10)}
 \def\y{6}\multiput(5,\y)(1,0){1}{\bp} \multiput(6,\y)(1,0){2}{\nat} \multiput(8,\y)(1,0){7}{\bn} \put(16,\y){\bn}
 \def\y{7}\multiput(5,\y)(1,0){2}{\bp} \multiput(7,\y)(1,0){2}{\nat} \multiput(9,\y)(1,0){8}{\bn}
 \def\y{8}\multiput(5,\y)(1,0){3}{\bp} \multiput(8,\y)(1,0){3}{\nat} \multiput(11,\y)(1,0){6}{\bn}
 \def\y{9}\multiput(8,\y)(1,0){2}{\bp} \put(10,\y){\nat} \multiput(11,\y)(1,0){8}{\bn} \multiput(20,\y)(2,0){1}{\bn}
 \def\y{10}\multiput(7,\y)(1,0){3}{\bp} \multiput(10,\y)(1,0){3}{\nat} \multiput(13,\y)(1,0){8}{\bn}
 \def\y{11}\multiput(9,\y)(1,0){3}{\bp} \put(12,\y){\nat} \multiput(13,\y)(1,0){8}{\bn}
 \def\y{12}\multiput(8,\y)(1,0){4}{\bp} \multiput(12,\y)(1,0){3}{\nat} \multiput(15,\y)(1,0){6}{\bn}
 \def\y{13}\multiput(9,\y)(1,0){5}{\bp} \put(14,\y){\nat} \multiput(15,\y)(1,0){6}{\bn}
 \def\y{14}\multiput(8,\y)(1,0){6}{\bp} \multiput(14,\y)(1,0){3}{\nat} \multiput(17,\y)(1,0){4}{\bn}
 {#3}
 \end{picture}}

 \newcommand{\sfbf}{\upshape\bfseries\sffamily}
 
 \newcommand{\YES}{\textcolor{Green}{\ding{52}}}
 \newcommand{\NO}{\textcolor{Red}{\ding{56}}}
 \newcommand{\Put}[3]{\put(#1,#2){\makebox(0,0){#3}}}
 \newcommand{\PICfig}[3]{\setlength{\unitlength}{#2mm}\begin{picture}(100,100)\Put{50}{50}{\psfig{#1}}{#3}\end{picture}}
 
  \definecolor{LightGray}{rgb}{.75,.75,.75}
\newcommand{\edit}[1]{#1}

\begin{document}
\onecolumn
\title{The extremes of neutron richness}
\author{F.~Miguel Marqu\'es\inst{1}
}                     


\institute{LPC Caen, Normandie Universit\'e, ENSICAEN, Universit\'e de Caen, CNRS/IN2P3, 14050 Caen, France}
\date{Received: date / Revised version: date}

%
\abstract{
A neutron star is pictured as a gigantic nucleus overwhelmed by the number of neutrons, unlike real atomic nuclei, that have a similar number of neutrons and protons. Is this true? What if we could find, or create nuclei without protons? How far can we go in neutron richness? Our common sense tells us that these neutral nuclei should not exist, but if they do they would change our knowledge on neutron stars, on the properties of nuclei in general, and ultimately on the nucleon-nucleon interaction itself, the building block of matter.
 This huge potential impact has pushed some \edit{ambitious} nuclear physicists to search for them since the 1960s. The first positive hints appeared only in the XXI century, and nowadays several collaborations are trying to corner these weird objects and give a definite answer to this crucial question. In this \edit{review} we will go through this fascinating quest, that started with humble experiments and has now reached a stage of ambitious and sophisticated projects, both in experiment and theory.
%
} 
\maketitle

\today

\bigskip \setcounter{tocdepth}{2} \tableofcontents \bigskip

\section{Nuclei without protons?} \label{s:introduction}

 The atomic nucleus has a relatively simple content: a given number of protons $Z$ and of neutrons $N$. These $A=Z+N$ ``nucleons'', however, may generate a huge variety of complex structures for certain $(Z,N)$ combinations.
 The basic link of these structures, the nucleon-nucleon force, remains qualitatively unknown, and only quantitative fits are available since the 1990s \cite{Machleidt_2001}.
 Although these fits are basically phenomenological and contain more than 50~parameters, they describe well the world database on $pp$ and $np$ scattering.
 However, already when attempting to model the very lightest nuclei, $^3$H and $^{3,4}$He, one must add a three-nucleon force. Only very recently, the three-nucleon force can be understood more quantitatively,
 for instance in the framework of chiral effective field theory \cite{Epelbaum_2020} by incorporating recent nucleon-deuteron scattering data \cite{Sekiguchi_2002,Sekiguchi_2004}.

 It is therefore surprising that a very simple model like the liquid drop, with only five parameters, provides an overall good description (within few MeV) of the energy of most of these complex nuclei. In the liquid drop model the binding energy of a nucleus, defined as the mass loss of the system $Zm_p+Nm_n-M(Z,N)$, has the form:
 \begin{equation} \label{e:drop}
 B(Z,N)\ =\ a_vA -a_sA^{2/3} -a_c\frac{Z^2}{A^{1/3}}
                 -a_a\frac{(N-Z)^2}{A} \pm\frac{\delta}{A^{1/2}}
 \end{equation}
 From left to right, the first three terms are justified by some nuclear properties analog to liquid drops: constant density and saturation (volume term), surface tension (surface term), and electrical repulsion between protons (Coulomb term).
 The other two terms are added in order to mimic some quantum properties: the Pauli principle favors configurations where $N\approx Z$, which fill the neutron and proton potential wells up to a similar energy (asymmetry term), and protons and neutrons couple to form pairs (pairing term) \cite{Povh_1999}.
 
 Even if we know that the nucleus is a much more complex object, the simplicity of Eq.~(\ref{e:drop}) leads us to `see' nuclei as liquid drops. We imagine them to have sharp limits, uniform volume, and homogeneous proton/neutron mixtures.
 In fact, most of the $(Z,N)$ combinations that lead to stable nuclei, the more or less 300 isotopes that form the so-called valley of stability, do exhibit those three properties. They have a radial `border', $R\approx r_0A^{1/3}$, and within this border the density is almost constant, $\rho(r)\approx\rho_0$, and the nucleon mixture is quite homogeneous, $\rho_n(r)\approx\frac{N}{Z}\rho_p(r)$.
 Eq.~(\ref{e:drop}) can be extended and refined to take into account other effects, like deformation, shell closures, Wigner term... but all of them are compatible with the liquid drop picture.
 The fact that about 3000 exotic nuclei, with $N/Z$ ratios different from those of the stable isotopes, have been studied in the past decades was neither in contradiction with it, as their lower binding was accounted for by the asymmetry term in Eq.~(\ref{e:drop}), or slightly more sophisticated versions.

 Only a few of these thousands of $(Z,N)$ combinations are in contrast with the liquid drop picture. For example, when the binding energy of some valence neutrons becomes very weak their wave function extends well beyond the nuclear core forming a halo, as in the ground states of $^{6}$He, $^{11}$Li and $^{14}$Be \edit{\cite{Haloes}}.
 In a few other cases nucleons cluster into $\alpha$ particles and the excess neutrons play the role of electrons in molecules, forming nuclear molecular states, as in the excited states of neutron-rich beryllium isotopes \edit{\cite{Molecules}}.
 Therefore, even if it is rare, we know that several neutrons can virtually `escape' from the nucleus to form a halo, and in some cases they can associate to other `fugitive' \edit{neutrons} to bind nuclear molecules.
 In those cases, however, we deal with multineutron systems that exist within a given frame, the nucleus. A question naturally arises: how would these neutron systems behave in the absence of witnesses (the core in halo nuclei or the $\alpha$ particles in nuclear molecules)?
 Can we build a nucleus without protons?
 
\begin{figure}[ht] \begin{center} \medskip
 \NZ{4.5}{10}{
  \Put{22.5}{2.5}{\psfig{file=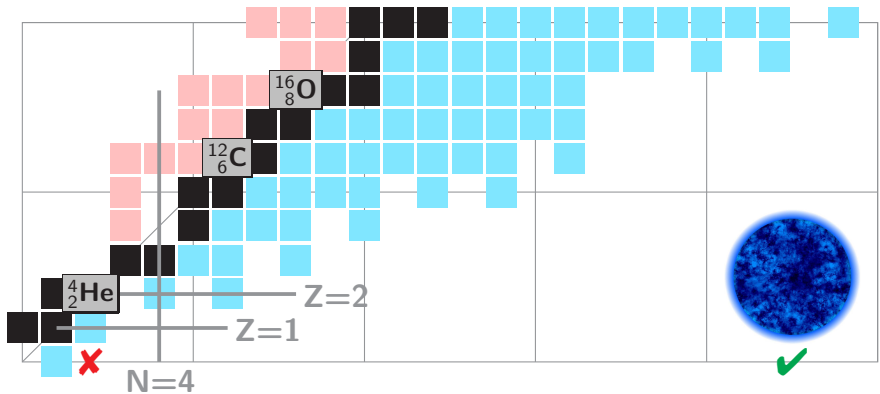,height=18mm}}\Put{2}{0}{\Large\NO}\Put{22.5}{0}{\Large\YES}\linethickness{.5mm}\Gray
  {\Line(1,1)(6,1)\Line(2,2)(8,2)\Line(4,8)(4,0)
   \large\sfbf\put(6.2,.6){Z=1}\put(8.2,1.6){Z=2}\put(3.,-.9){N=4}}
  \def\sizeISO{\normalsize}
  \Put{2}{2}{\ISO{2}{4}{He}{LightGray}}
  \Put{6}{6}{\ISO{~6}{12}{C}{LightGray}}
  \Put{8}{8}{\ISO{~8}{16}{O}{LightGray}}} \bigskip
 \end{center}
 \caption{The chart of nuclei up to $Z=10$ (neon). The 
 black boxes represent the stable isotopes (some examples are shown in gray). For the bottom row of the chart ($Z=0$), the only evidence is that the dineutron ($N=2$) is unbound and that a neutron star ($N\sim10^{57}$, \edit{of course} not to scale) is bound. The gray lines show the location of the $N=4$ and $Z=1,2$ nuclei, that will be discussed in Fig.~\ref{f:masses}.}\label{f:chart}
\end{figure}
 
\begin{figure}[ht] \begin{center}
 \psfig{file=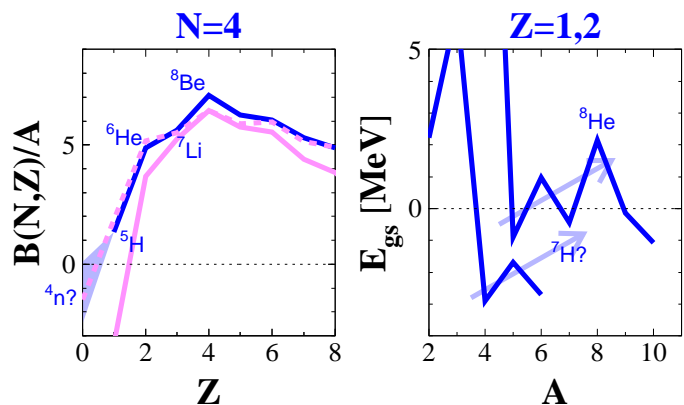,width=9.5cm}
 \end{center}
 \caption{The blue lines correspond to the experimental binding energy per nucleon (left) for $N=4$ isotones and the ground state energy (right) for hydrogen \edit{(bottom)} and helium \edit{(top)} isotopes. The experimental values for $^4$n and $^7$H are not well known yet, and the light blue area/arrows represent possible trends.
 The pink solid line on the left panel corresponds to Eq.~(\ref{e:drop}) for a standard set of parameters, and the dashed line to a simple modification of its asymmetry term considering surface effects.}\label{f:masses}
\end{figure}
 
 The present experimental answer is clear: we cannot. The lightest candidate, the dineutron, is already unbound, and in order to find a bound system of neutrons we need to build something much bigger than a nucleus, a neutron star, as sketched in Fig.~\ref{f:chart}.
 But let us play a very easy exercise and set $Z=0$ in Eq.~(\ref{e:drop}). For a typical set of parameters \cite{Povh_1999}, the result is that multineutrons should be unbound by about 15~MeV/neutron (the tetraneutron in particular by about 70~MeV).
 However, we can see in the left panel of Fig.~\ref{f:masses} that the standard liquid drop overestimates the effect of asymmetry and diverges from experiment for increasing $N/Z$.
 Some versions of the liquid drop formula have included surface effects in the calculation of $a_a$, since the effect of the asymmetry term decreases for increasing surface-to-volume ratio (lighter nuclei) \cite[p.~197]{Hornyak_1975}. These liquid drop formulae provide a much better description of very asymmetric light nuclei. An example of those is the dashed line in Fig.~\ref{f:masses}, and from that point of view the question of a bound or near-threshold resonant state in the $4n$ system may deserve some thought.

 Although the experimental value for $^4$n is missing, the binding energies per nucleon for light $N=4$ isotones follow a decreasing sequence $^8$Be-$^7$Li-$^6$He-$^5$H modulated by the $\pm\delta$ pairing term of Eq.~(\ref{e:drop}). The next member of the sequence should continue the decrease, but with a $+\delta$ contribution. The light blue area in Fig.~\ref{f:masses} shows that there is room for a tetraneutron state around threshold.
  Note that the neighboring $^5$H is unbound but has a positive binding energy of about 6~MeV: the $4n+p$ system is 6~MeV lighter than the \edit{free} five nucleons. $^5$H is unbound only due to the strong binding of the triton (7.7~MeV), which induces an instantaneous decay into $^3$H$+2n$.
  But in the case of $^4$n there are no bound subsystems, and then even a 1~eV binding energy would lead to a bound tetraneutron.

 Another clue can be extracted from the known masses of hydrogen and helium isotopes, on the right of Fig.~\ref{f:masses}.
 The overall trend for all the elements is that the binding of the ground state (the distance to the first particle threshold) decreases monotonically, besides the odd-even staggering, as more neutrons are added to the most stable $(Z,N)$ combination.
 Hydrogen and helium, however, are exceptions to this rule, as shown by the light blue arrows. It is intriguing that this extra binding provided by additional neutrons appears only for nuclei with very few protons, and that the maximum effect seems to be associated to four neutrons: $\alpha+4n$ leads to the particularly stable $^8$He\footnote
 {Its $4n$ separation energy $S_{4n}(^8$He$)=3.1$~MeV has been the upper limit of the tetraneutron binding energy for decades, since a higher value would make $^8$He unbound. Recent mass measurements have decreased it to $S_{4n}(^{19}$B$)=1.5(4)$~MeV \cite{AME2016}.},
 and $^5$He$+4n$ leads to an almost bound $^9$He. In this respect, as we will see, the mass of $^7$H is of capital importance.

\begin{figure}[ht] \begin{center}
  \psfig{file=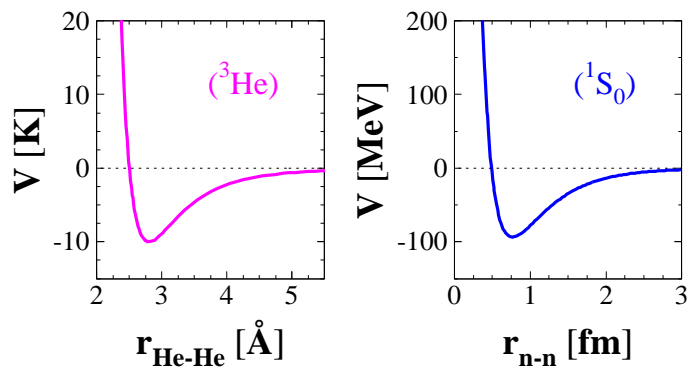,width=9.5cm}
 \end{center}
 \caption{Comparison between the $^3$He inter-atomic potential \edit{\cite{He3}} and the $nn$ potential in the singlet state \edit{\cite{Vnn}}.}\label{f:3He}
\end{figure}

 Finally, there is an analog case in atomic physics, liquid ($^3$He)$_{\rm N}$ drops. Indeed, $^3$He atoms are fermions, like the neutron, and their attractive interaction is also too weak to form a dimer, as in the dineutron case.
 Since a very high number of atoms form a liquid drop, theoretical studies have been undertaken to find the critical number of atoms needed to form a bound system, leading to N$\,\sim30$ \cite{Guardiola_PRL84_2001}.
 Could it be that a critical number exists also for neutron drops? The interaction potentials, although at very different scales, have indeed a similar shape (Fig.~\ref{f:3He}). The same type of calculations are, however, not yet available at the nuclear level, as the $nn$ potential is much more complex than just the central part drawn in Fig.~\ref{f:3He}.

 In any event, the discovery of such neutral systems as bound or resonant states would certainly have far-reaching implications for the modeling of neutron stars \cite{4n_stars_USAL_EJPA155_2019}, and most importantly for our understanding of the properties of nuclei in general and the underlying nucleon-nucleon interaction itself, the building block of matter.
 In the following I will review the different problems that the quest for neutral nuclei has been facing since the early 1960s \cite{Ogloblin89}.  
 The first positive hints appeared only in the XXI century \cite{Marques02_4n_recoil,Kisamori16_4n_DCX}, and nowadays several collaborations are trying to corner these weird objects and give a definite answer to this crucial question, both in experiment and theory.

\section{Production and detection of neutral nuclei} \label{s:production}

\def\longarrow{\relbar\!\!\relbar\!\!\relbar\!\!\relbar\!\!\relbar\!\!\relbar\!\!\rightarrow}
\def\hof{\hskip5.5ex}\def\inter{\\[-1.55ex]\hof$\Downarrow$\\[-.55ex]\hof$\Uparrow$\\[-1.6ex]}
\newcommand{\scheme}[3]{\begin{tabular}{l}#1\inter#2\\[0mm]\multicolumn{1}{c}{\normalsize#3}\end{tabular}}

\begin{figure}[ht] \begin{center}
  \psfig{file=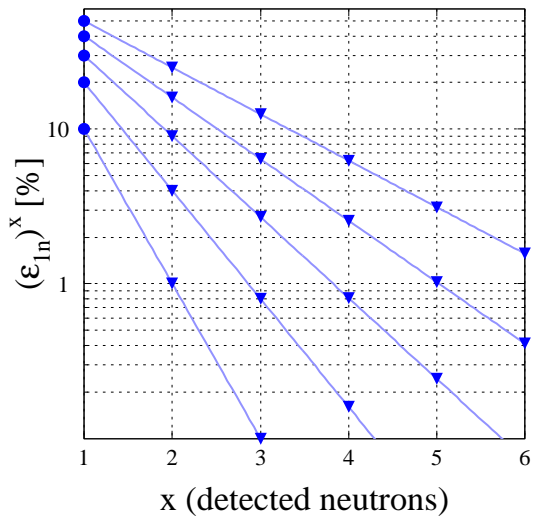,width=7cm}
 \end{center} \caption{The neutron efficiency to the power of the number of detected neutrons, as a function of the latter. From bottom to top, the lines/symbols correspond to neutron efficiencies $\varepsilon_{1n}$ of 10, 20, 30, 40 and 50\%, respectively. The vertical scale goes from 1\permil\ to 60\%.
 In practice, due to cross-talk effects \cite{cross-talk} the multineutron ($x>1$) efficiency will be even lower, $\varepsilon_{xn}<(\varepsilon_{1n})^x$.} \label{f:efficiency}       
\end{figure}

 Detecting neutral particles represents an experimental challenge in general. Charged particles crossing a detector material interact with the atomic electron clouds, leading to detection efficiencies close to 100\%. Neutral particles, however, must interact with a nucleus, as Chadwick realized in his discovery of the neutron \cite{Chadwick32}. Since this is a much less probable process, the typical detection efficiencies (for similarly sized detectors) are of the order of few per cent.
 Moreover, the detection of $x$ neutrons becomes exponentially harder, since the efficiency will decrease roughly as $\varepsilon_{xn}\approx(\varepsilon_{1n})^{\,x}$, as shown in Fig.~\ref{f:efficiency}.

 To make things worse, the latter is just an upper limit due to the effect of cross-talk: one neutron may scatter on several nuclei in the detector array and therefore be detected several times. The identification of these events, that mimic the detection of several independent neutrons, requires the application of rejection algorithms with subsequent losses of total efficiency \cite{cross-talk}. 

\begin{figure}[ht]
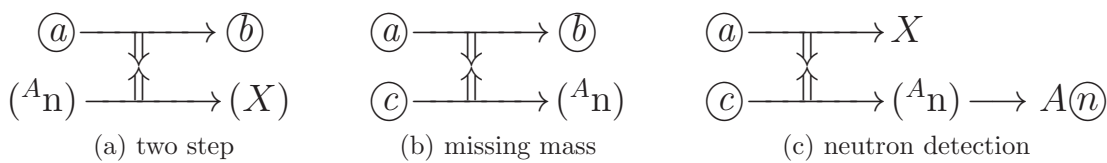
 \begin{center}
 {\Large \scheme{\textcircled{\it{a}}$\longarrow\,$\textcircled{\it{b}}}
        {\hskip-1.8ex$(^A\mbox{n})\!\longarrow\!(X)$}{(a) two step}
 \hskip2.ex \scheme{\textcircled{\it{a}}$\longarrow\,$\textcircled{\it{b}}}
        {\textcircled{\it{c}}$\longarrow\!(^A\mbox{n})$}{(b) missing mass}
 \hskip2.ex \scheme{\textcircled{\it{a}}$\longarrow\!X$}
        {\textcircled{\it{c}}$\longarrow\!(^A\mbox{n})\!\longrightarrow A$\textcircled{$n$}}{(c) neutron detection}
 }\end{center}
 \caption{The three main categories in which all experiments searching for $^A$n multineutrons can be classified.} \label{f:categories}       
\end{figure}

 Even if one obviates the detection problem, the fact that neutrons are difficult to guide and unstable themselves (half-life of about 10~min) does not allow us to build a system of several neutrons out of its components.
 Therefore, all the experimental approaches must face the challenge of overcoming those two issues: how to build multineutrons and then how to detect them.
 Paradoxically, even if the aim of these experiments is the `observation' of systems of several neutrons, almost all of them have in common the absence of any neutron detection, due to the aforementioned very low $\varepsilon_{xn}$ efficiencies. 
 According to the principle they use, we can identify three categories of experiments that are schematically shown in Fig.~\ref{f:categories}.

 In Fig.~\ref{f:categories}a, representing the two-step category, a bound multineutron is supposed to be produced in a previous reaction, and then one signs its subsequent interaction with a nucleus \textcircled{\it{a}} that induces a transformation into \textcircled{\it{b}}. The main advantage of this category is that it only requires the detection of one charged particle. However, there are many related problems. Only bound $^A$n states can lead to this second reaction, there is no sensitivity to the multineutron energy, and only a lower limit of the mass number $A$ can be inferred from the transformation into \textcircled{\it{b}}.
 Moreover, a contaminant different from \textcircled{\it{a}} could lead to \textcircled{\it{b}} without requiring a multineutron, and the generally uncontrolled previous reaction may produce a huge background of many particles, that could be eventually responsible also for the production of \textcircled{\it{b}}.

 In Fig.~\ref{f:categories}b, representing the missing-mass category, the multineutron is supposed to be formed in a two-body collision \textcircled{\it{a}}+\textcircled{\it{c}} leading to a two-body final state \textcircled{\it{b}}+$^A$n.
 Only in that case, the detection of \textcircled{\it{b}} can sign the population of states in the missing multineutron system through the unique constraints of two-body kinematics. This technique requires also the detection of only one charged particle, and has two additional advantages: the multineutron mass number $A$ is well defined, and both bound and resonant $^A$n states can be probed.
  However, it still has issues. The cross-sections that bring all the protons into \textcircled{\it{b}} without breaking it are generally (extremely) low, and any beam or target contaminant different from \textcircled{\it{a}} or \textcircled{\it{c}} would lead to a missing partner(s) of \textcircled{\it{b}} that is not a multineutron.
 
 In Fig.~\ref{f:categories}c, representing the detection of all the neutrons, a multineutron state is supposed to be formed in a given reaction and the different neutrons from its decay (if resonant) or from its breakup (if weakly bound) would be detected. It seems the most unambiguous, and it carries all the information about the state: mass number, energy, and even internal correlations. However, Fig.~\ref{f:efficiency} shows the inherent difficulty of this technique, and explains why it has not been used in the past.
 
 There have been several tens of experiments since the early 1960s, and in the following I will try to review them from the perspective of these three categories.

\subsection{Two steps: activation reactions}

 In this category, illustrated in Fig.~\ref{f:categories}a, the multineutron is supposed to be produced in a first step, high-flux reaction, not necessarily well-characterized, like spallation or induced fission. Assuming that a bound multineutron was produced, together with many other particles, we may use it in a secondary two-body reaction that transforms a given sample in a unique way.
 Therefore, we need to demonstrate that the sample was transformed, but above all that it could not be transformed in any other alternative way.

\begin{figure}[ht]
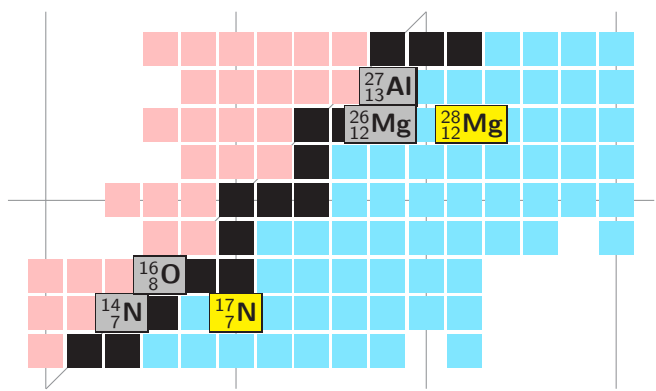
 \begin{center}
 \NZcropNMg{5.}{}{
  \def\sizeISO{\normalsize}
  \Put{7}{7}{\ISO{~7}{14}{N}{LightGray}}
  \Put{8}{8}{\ISO{~8}{16}{O}{LightGray}}
  \Put{14}{13}{\ISO{13}{27}{Al}{LightGray}}
  \Put{13.8}{12}{\ISO{12}{26}{Mg}{LightGray}} \def\sizeISO{\normalsize}
   \Put{10}{7}{\ISO{~7}{17}{N}{Yellow}}
   \Put{16.2}{12}{\ISO{12}{28}{Mg}{Yellow}}}
 \end{center}
 \caption{Zoom on the chart of nuclei for $Z=5$--15 and $N=5$--20. In the 1960s, Schiffer \cite{Schiffer63_4n_nU} and Cierjacks \cite{Cierjacks65_4n_d238U} tried to transform natural isotopes of nitrogen, oxygen, magnesium and aluminum into unstable $^{17}$N and $^{28}$Mg, through the absorption of two or three neutrons from an hypothetical tetraneutron. No decays of those two isotopes were observed.}\label{f:decay}
\end{figure}

 As early as 1963, Schiffer \etal\ irradiated samples of nitrogen and aluminum in a nuclear reactor \cite{Schiffer63_4n_nU}. They were looking for the production of bound tetraneutrons in the fission of the uranium fuel. If they had been produced, they might have observed the reactions $^{14}$N$(^4$n$,n)^{17}$N and $^{27}$Al$(^4$n$,t)^{28}$Mg in their samples (see Fig.~\ref{f:decay}).
 These two isotopes have a characteristic decay: $^{17}$N emits a neutron after its $\beta$ decay into $^{17}$O, and $^{28}$Mg decays into $^{28}$Al and both of them emit $\gamma$-rays. However, they could not observe the corresponding neutrons or $\gamma$-rays above the different backgrounds.
 Two years later, Cierjacks \etal\ carried out a similar experiment. Instead of a reactor, they induced the first reaction by bombarding a uranium target with deuterons, and placed samples of nitrogen, oxygen, magnesium and other heavier elements around the uranium target \cite{Cierjacks65_4n_d238U}. For the three lighter samples, they searched for bound tetraneutrons emitted in uranium fission through the reactions $^{14}$N$(^4$n$,n)^{17}$N, $^{16}$O$(^4$n$,t)^{17}$N and $^{26}$Mg$(^4$n$,2n)^{28}$Mg (see Fig.~\ref{f:decay}). As Schiffer, they were not able to observe the corresponding neutrons or $\gamma$-rays above the background.

\begin{figure}[ht] \begin{center}
  \psfig{file=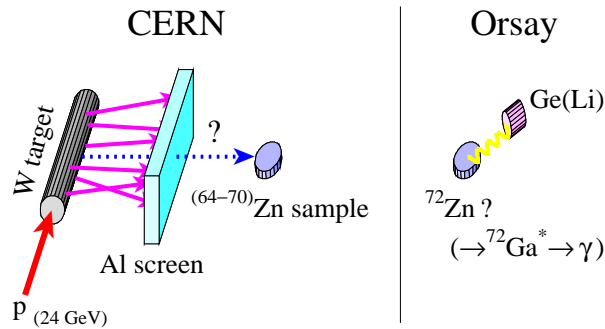,width=8.5cm}
 \end{center} \caption{Schematic view of D\'etraz's experiment \cite{Detraz77_68n_pW}. In the collision of protons and tungsten, a neutral cluster was supposed to be produced, go through an aluminum screen, and induce $(^A$n$,xn)$ reactions on a natural zinc sample. The new radioactive $^{72}$Zn isotope was detected in another laboratory through its decay into $^{72}$Ga$^*$ with a germanium detector.} \label{f:Detraz}       
\end{figure}
 
 Surprisingly, in 1977 an experiment at CERN following a similar principle obtained a clear signal of bound multineutrons. D\'etraz exposed a natural zinc sample to a tungsten block irradiated with a proton beam of 24~GeV \cite{Detraz77_68n_pW}. As we can see in Fig.~\ref{f:Detraz}, block and sample were separated by an aluminum screen, supposed to be thick enough to stop any charged particle.
 However, any neutral cluster that might have been produced in the collision would have been able to go through the screen and deposit several neutrons (at least two) on the natural zinc isotopes through the reaction $^{64,66,67,68,70}$Zn$(^A$n$,xn)^{72}$Zn. 
 The relatively long half-life of $^{72}$Zn (46.5~h) and its daughter $^{72}$Ga (14.1~h), and the several $\gamma$-rays emitted by the latter, made it possible to remove the sample from the high-activity area and perform a clean measurement of those $\gamma$-rays.
 After several-day exposure, the 100~g of natural zinc were sent by air freight to Orsay, where a germanium detector revealed the unambiguous production of $^{72}$Zn through five prominent $\gamma$-rays of its daughter $^{72}$Ga.
 The production of $^{72}$Zn could have been the result of any ($A>1$) bound multineutron, but since the previous searches for $A=2$--4 had failed, D\'etraz concluded that most likely bound hexa- or octaneutrons ($A=6,8$) had been produced \cite{Detraz77_68n_pW}.
 
 A few months later, Turkevich \etal\ tried to confirm D\'etraz's exciting results using a similar principle \cite{Turkevich77_6n_pU}. The setup was similar to the one in Fig.~\ref{f:Detraz}, but the target was uranium, the proton energy was 700~MeV, and the sample was a lead block. 
 The transformation searched for in the sample was the absorption of $4n$ belonging to an $A\geqslant4$ multineutron by a natural lead nucleus, following the reaction $^{208}$Pb$(^A$n$,xn)^{212}$Pb.
 If $^{212}$Pb had been produced, they would have seen it through the $\alpha$ particles \edit{emitted by the daughter of its $\beta$-decay chain, 
 $^{212}$Po}.
 No trace of what they called ``polyneutrons'' was found, shedding doubt on D\'etraz's results.
 In 1980, de Boer \etal\ persisted.
 They irradiated a block of tellurium with $^3$He ions and searched for the interaction of bound tetraneutrons in the same block through the $^{130}$Te$(^4$n$,2n)^{132}$Te reaction \cite{deBoer80_46n_3He130Te}.
 No $\gamma$-rays from the decay of $^{132}$Te were observed, and de Boer concluded, \edit{with the agreement of D\' etraz}, that the latter had in fact underestimated the transmission of tritons through his aluminum screen and that the reaction observed had been $^{\rm(nat)}$Zn$(t,p)^{72}$Zn, without any need of multineutrons \cite{deBoer80_46n_3He130Te}.

 Following these results, the two-step or activation probe was abandoned for more than thirty years in favor of the cleaner missing-mass experiments.
 Anyhow, in 2012 Novatsky \etal\ revisited the technique by inducing uranium fission with 62~MeV $\alpha$ particles, and then searched for the interaction of bound multineutrons within strontium and aluminum samples shielded by thin Kapton and beryllium screens \cite{Novatsky12_6n_a238U,Novatsky13_6n_a238U}.
 The authors claimed they had observed the reaction $^{88}$Sr$(^A$n$,xn)^{92}$Sr, through $\gamma$-rays from the daughter $^{92}$Y \cite{Novatsky12_6n_a238U}, and the reaction $^{27}$Al$(^A$n$,p\,xn)^{28}$Mg, through $\gamma$-rays from the decay into $^{28}$Al and $^{28}$Si \cite{Novatsky13_6n_a238U}.
 They concluded that, due to the lack of evidence for a bound tetraneutron, their results should ``certainly'' correspond to heavier ($A\geqslant6$) clusters.
 However, the experimental hall exhibited strong activity, the number of channels following fission is very high, and it could be that the screens that were used were not thick enough to shield all the many charged particles produced, as it had been the case for D\'etraz in 1977.
 
 To conclude, the so-called activation probe appeared as a powerful tool from the very first years, lead to some hope about bound multineutrons in the late 70s, but after the refutation of the claim it was left out and is not being followed nowadays in any large-scale facility.

\subsection{Missing mass: pion beams}

 In this category, illustrated in Fig.~\ref{f:categories}b, the multineutron is supposed to be produced in a two-body reaction. Due to energy and momentum conservation, the observation of peaks in the energy spectrum of the final-state partner would reveal multineutron states, bound or resonant.
 A very clean example is the double charge exchange (DCX) reaction $(\pi^-,\pi^+)$, since pions do not have excited states. The incoming negative pion becomes positive, changing two protons in the target into neutrons. By measuring the momenta of the pions on a $^{3,4}$He target, one can measure the missing mass of the $^{3,4}$n system. Obviously, since no other helium isotopes can be used as targets, no other multineutrons are accessible.
 
 In 1965 Gilly \etal\ started this axis searching for the tetraneutron with the $^4$He$(\pi^-,\pi^+)4n$ reaction \cite{Gilly65_4n_pi4He}. However, they found no peaks in the $\pi^+$ spectrum, only a continuum that could be explained by introducing a final-state interaction (FSI) between two of the neutrons.
 Five years later Sperinde \etal\ extended this search to the trineutron using a $^3$He target with the $^3$He$(\pi^-,\pi^+)3n$ reaction \cite{Sperinde70_3n_pi3He}. As Gilly, no peaks were observed, but an enhancement at low $3n$ energies suggested a possible $3n$ resonance. They confirmed this enhancement in 1974 \cite{Sperinde74_3n_pi3He}, but it was finally explained by FSI effects between the neutrons \cite{Jibuti85_34n_pi34He}.
 
\begin{figure}[ht] \begin{center}
  \psfig{file=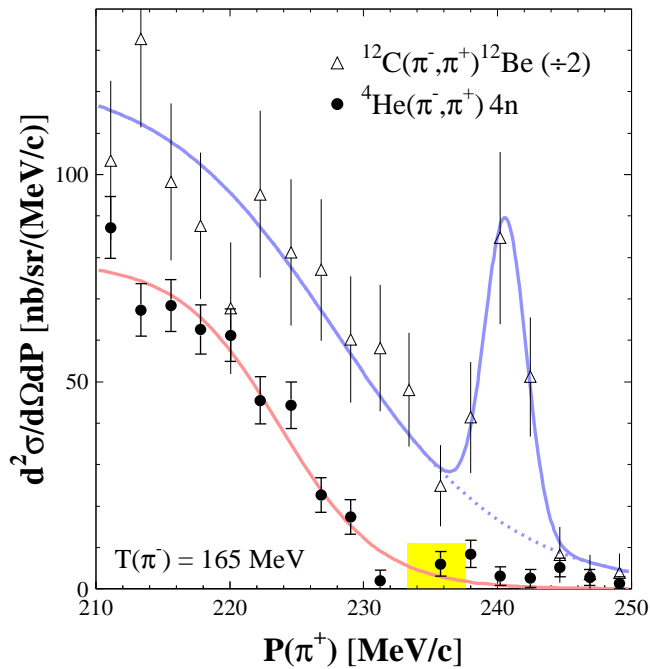,width=8.5cm}
 \end{center} \caption{Experimental results for the reactions $^{12}$C$(\pi^-,\pi^+)^{12}$Be (triangles, divided by 2) and $^4$He$(\pi^-,\pi^+)4n$ (circles). The curves are fits to guide the eye, with a Woods-Saxon distribution only (red) plus an additional Gaussian function (blue). The peak in the $^{12}$C channel corresponds to the formation of the $^{12}$Be ground and two first excited states, and the range in yellow in the $^4$He channel to the region expected for a bound tetraneutron. Adapted from \cite{Ungar84_4n_pi4He}.} \label{f:Ungar}       
\end{figure}
 
 In 1984 Ungar \etal\ revisited the tetraneutron with the $^4$He$(\pi^-,\pi^+)^4$n reaction \cite{Ungar84_4n_pi4He}.
 In order to cross-check their technique, they measured first a similar reaction with known bound final states, $^{12}$C$(\pi^-,\pi^+)^{12}$Be. As can be clearly seen in Fig.~\ref{f:Ungar} (triangles), the $\pi^+$ spectrum exhibited a prominent peak around the threshold that reflected the formation of bound states in $^{12}$Be.
 When they changed the target to $^4$He (the $4n$ channel), however, no apparent peak was observed (Fig.~\ref{f:Ungar}, circles).
 The expected region for a bound $^4$n, defined by the aforementioned $S_{4n}(^8$He) and shown in yellow, contained a few events. However, their rate was similar to the events above this region, which is kinematically forbidden, and they were thus associated to the background due to imperfect $\pi^+$ detection in the spectrometer.
 Two years later, Stetz \etal\ extended the search to the trineutron with the reactions $^{3,4}$He$(\pi^-,\pi^+)^{3,4}$n \cite{Stetz86_34n_pi34He}, at several energies and angles, but again they found no evidence of trineutron or tetraneutron states.
 
 In 1989 Gorringe \etal\ repeated Ungar's experiment at lower energy and higher angle \cite{Gorringe89_4n_pi4He}. While they also observed a few events in the possible bound $^4$n window, they found them consistent with the estimated continuum contribution, and only an upper limit of the production cross-section of an hypothetical tetraneutron could be deduced.
 At the end of the 90s, Yuly \etal\ \cite{Yuly97_3n_pi3He} and Gr\"ater \etal\ \cite{Grater99_3n_pi3He} undertook a systematic study of the $^3$He$(\pi^-,\pi^+)3n$ reaction at 65--240~MeV, and found no evidence for trineutron states.
 These experiments closed an extensive research program, from 1965 to 1999, on the $^{3,4}$He$(\pi^-,\pi^+)^{3,4}$n reactions. With the closing of the XX century, it seemed that the lightest multineutrons did not exist.

 But there had been other experiments using negative pion beams.
 Already in 1976 Bistirlich \etal\ studied the single charge exchange on hydrogen, $^3$H$(\pi^-,\gamma)3n$ \cite{Bistirlich76_3n_pi3H}, but they found no evidence for a trineutron in the $\gamma$ spectrum. Four years later, Miller \etal\ repeated the experiment with better statistics and resolution \cite{Miller80_3n_pi3H}, and confirmed the results.
 In parallel, Chultem \etal\ proposed an original use of a $\pi^-$ beam in a two-step process \cite{Chultem79_4n_pi208Pb}. If a bound tetraneutron had been produced in the $^{208}$Pb$(\pi^-,\pi^+)^4$n reaction, by DCX on an $\alpha$ cluster inside lead, they might measure the tetraneutron absorption by another lead nucleus transforming it into $^{212}$Pb, a probe similar to Turkevich's \cite{Turkevich77_6n_pU}. But as the latter, they found no $\alpha$ particles from its decay chain. 
 Finally, in 1991 Gornov \etal\ studied the $3n$ missing-mass spectrum in the reactions $^9$Be$(\pi^-,t\,^3$He$)$ and $^9$Be$(\pi^-,d\,^4$He$)$ \cite{Gornov91_3n_pi9Be}. It was tentatively described using a very broad trineutron resonance at 3~MeV, but the very limited resolution did not allow to draw firm conclusions. 

 In summary, after more than thirty years of experiments with pion beams, only upper limits following the non-observation of multineutrons have been set, and the technique is not being presently used.

\subsection{Missing mass: transfer reactions}

 This category still corresponds to Fig.~\ref{f:categories}b, the production of a multineutron in a two-body reaction, which would be thus revealed through peaks in the energy spectrum of its final-state partner.
 However, it does not rely on changing two protons into two neutrons, like the DCX reactions that limited the search to $^{3,4}$He targets.
 The multineutron system was probed in a transfer reaction between two stable nuclei, increasing the possible combinations.
 But the cross-sections were expected to be still very low, since one must: transfer all the protons away from one nucleus; sometimes bring neutrons back in the opposite sense; and in any event lead exclusively to a two-body final state.
 
 The first such experiment was carried out in 1965 by Ajda\v{c}i\'c \etal\ using a simple transfer reaction on the triton, $^3$H$(n,p)3n$ \cite{Ajdacic65_3n_n3H}. Some events were observed at a missing mass of about 1~MeV below the $3n$ threshold, the first candidates of a (quite) bound trineutron. However, knowing that the very first experiments had failed to find a bound tetraneutron \edit{(for example Ref.~\cite{Schiffer63_4n_nU})}, more likely to exist due to pairing, they concluded that their result was ``highly improbable''.
 One year later, Thornton \etal\ repeated the same experiment with better resolution \cite{Thornton66_3n_n3H}, and found no evidence for a bound trineutron.
 Two years later Ohlsen \etal\ used a triton beam and searched already for a more complex transfer reaction, $^3$H$(t,^3$He$)3n$ \cite{Ohlsen68_3n_t3H}. The missing mass reconstructed from $^3$He lead to a deviation from four-body phase space, only at forward angles, that could be consistent with a low-energy trineutron resonance. However, they were not able to exclude an effect from the reaction mechanism itself.
 
\begin{figure}[ht] \begin{center}
  \psfig{file=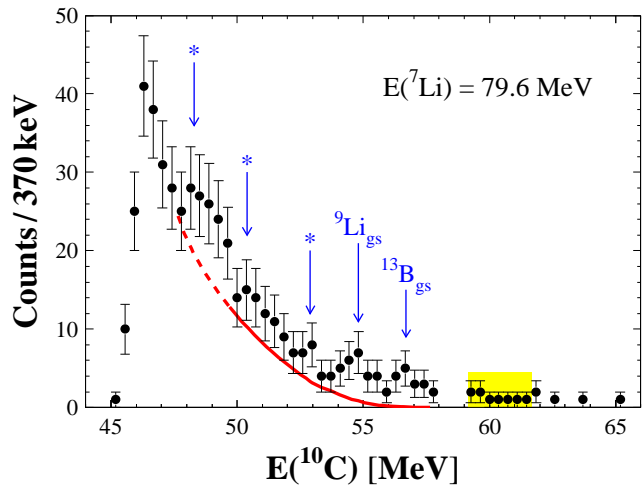,width=8.5cm}
 \end{center} \caption{Energy spectrum of $^{10}$C from the $^7$Li$(^7$Li$,^{10}$C$)4n$ reaction. The contributions from known contaminant reactions are indicated in blue either explicitly or with an asterisk. The red curve corresponds to five-body phase space, and the range in yellow to the region expected for a bound tetraneutron. Adapted from \cite{Cerny74_43n_transfer}.} \label{f:Cerny}       
\end{figure}

 Heavier nuclei started to be used in 1974, when Cerny \etal\ searched for the tri- and tetraneutron in the reactions $^7$Li$(^7$Li$,^{11}$C$)3n$ and $^7$Li$(^7$Li$,^{10}$C$)4n$ \cite{Cerny74_43n_transfer}. 
 Concerning the trineutron, the $3n$ missing-mass spectrum could be well described by four-body phase space, plus some small peaks from known target contaminants (that lead to $^{11}$C partners different from $3n$). 
 In the tetraneutron channel, however, the low $^{10}$C production led to a poor separation from the tail of the much stronger $^{11}$C distribution. The resulting $4n$ missing-mass spectrum could be described by five-body phase space plus the known contaminants, as can be seen in Fig.~\ref{f:Cerny}. Although some events were visible in the possible region for bound $^4$n, the signal was not significant with respect to the background level.

 In 1988 Belozyorov \etal\ improved on the main issues of Cerny's work, the target purity and fragment identification, and searched for the trineutron in the $^7$Li$(^{11}$B$,^{15}$O$)3n$ reaction and for the tetraneutron in the $^7$Li$(^{11}$B$,^{14}$O$)4n$, $^7$Li$(^9$Be$,^{12}$N$)4n$ and $^9$Be$(^9$Be$,^{14}$O$)4n$ reactions \cite{Belozyorov88_43n_transfer}.
 The missing-mass spectra above the $3n$ and $4n$ thresholds were well described by the corresponding four- and five-body phase space, showing no evidence for resonances, while few events below the $4n$ threshold were found to be consistent with the background.
 In 1995 Bohlen \etal\ used a $^{14}$C beam in order to probe the trineutron in the reaction $^2$H$(^{14}$C$,^{13}$N$)3n$ on a CD$_2$ target \cite{Bohlen95_3n_transfer}. The missing mass was fully described without trineutron resonances.
 Finally, in 2005 Aleksandrov \etal\ repeated Cerny's experiment, with similar beam energy and target, and obtained the same negative results for both the tri- and tetraneutron \cite{Aleksandrov05_43n_transfer}.
 
 This technique had appeared as a good compromise between pion DCX and activation. It could access heavy multineutrons from a variety of beam-target combinations, and the potential signals were supposed to be unambiguous. However, the absence in practice of clear signals led towards a need for higher purities, and the technique has been put aside for the last fifteen years.

\section{Positive signals in the XXI century}

 The experiments performed in the XX century used mainly stable beams and targets. The beams could thus be very intense, but building a neutral system from balanced combinations of protons and neutrons required reactions with very low cross-sections.
 At the dawn of the XXI century, the use of the neutron-rich beams that had been developed in the previous decade appeared as the natural next step in this field.
 Logically, the two positive signals arrived in the facilities that dominated the physics of exotic beams, first at GANIL in the 2000s and then at RIKEN in the 2010s.

\begin{figure}[ht]
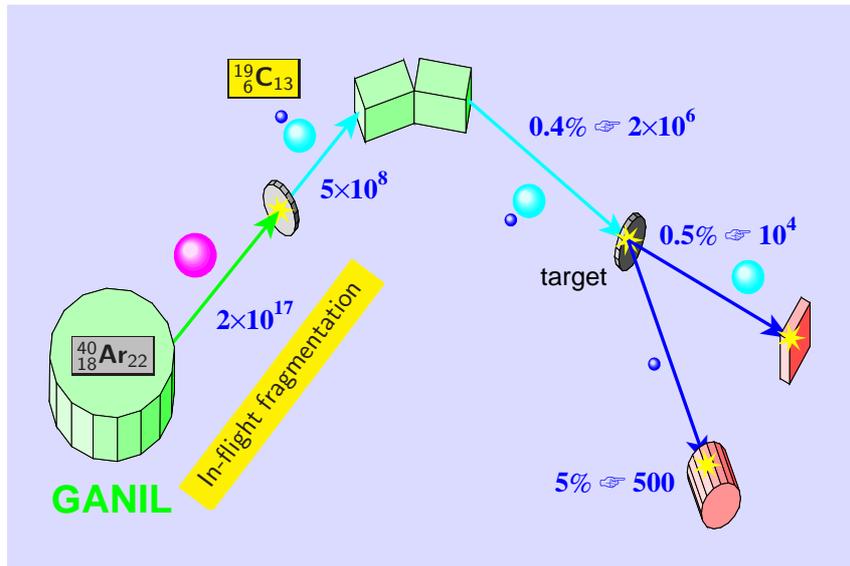
 \begin{center}
 \PICfig{file=jc2002_fra,height=7.5cm}{.75}
  {\normalsize\sf\Put{23}{33}{\rotatebox{53}{\colorbox{Yellow}{~In-flight fragmentation~}}}
  \def\sizeISO{\normalsize}
   \Put{-6.5}{38}{\ISO{18}{40}{Ar$_{\sf22}$}{LightGray}}
   \Put{20}{87}{\ISO{~6}{19}{C$_{\sf13}$}{Yellow}}}
 \end{center}
 \caption{Schematic illustration of the in-flight fragmentation technique using an experiment at the GANIL accelerator facility \cite{Marques_19C}. A primary argon beam was accelerated and fragmented in flight, producing among many other fragments $^{19}$C, with a half-life of only 46~ms. Once guided towards a secondary target in the experimental hall, the reaction studied was the breakup into $^{18}$C$+n$. The numbers in blue at each stage correspond to the approximated total number of events during the whole experiment.} \label{f:fragmentation}       
\end{figure}

 There are several techniques for the production of exotic beams, but the one that was used in both experiments was the in-flight fragmentation of a stable beam \edit{\cite{InFlight}}. I have illustrated its principle in Fig.~\ref{f:fragmentation} with another experiment that was carried out at GANIL, the breakup of a secondary beam of $^{19}$C \cite{Marques_19C}, with $(Z,N)=(6,13)$.
 One must accelerate a high-intensity stable beam with $(Z,N)\geqslant(6,13)$, in this case $^{40}$Ar, and break it up onto a primary production target. Due to the beam velocity all the fragments will be forward focused, with many different $(Z,N)\leqslant(18,22)$ combinations (the farther from $^{40}$Ar and/or the higher the $N/Z$ ratio, the lower the probability).
 Using magnetic elements, the desired combination is guided up to the experimental hall, where the second reaction takes place.
 
 The difficulty of these experiments can be understood from the approximated numbers in Fig.~\ref{f:fragmentation}. Out of the $2\times10^{17}$ argon ions accelerated by GANIL cyclotrons, only 500~million $^{19}$C ions were produced, a probability of the order of few in a billion. Due to the divergence of these ions, and to the strong selection filters used in order to suppress all the tails of the many more significant contributions from argon breakup, only 2~million $^{19}$C ions reached the secondary target. Despite a relatively thick target and the high cross-section, only 10,000 of them broke up. Finally, due to the small neutron detection efficiency, only 500 complete $^{18}$C$+n$ events were observed. Therefore, in this example there were fifteen orders of magnitude between the number of primary ions accelerated and of observed events.

\subsection{The GANIL result}

 In addition to the low cross-section \edit{of the reactions} that had been previously used, the potential multineutron signal often had shared parts of the spectra with background from contaminant species, and due to the low cross-sections, the background contributions had become too important for a signal to be clearly established.
 In 2002 Marqu\'es \etal\ proposed at the GANIL facility a new technique that could solve those issues \cite{Marques02_4n_recoil}.
 They considered the possible preformation of multineutrons inside very neutron-rich nuclei, similar to the preformation of $\alpha$ particles in the process of $\alpha$ decay.
 Within this scenario, the complex formation step of multineutrons \edit{used earlier} could be reduced to the breakup of one of those nuclei,
 with an increase in cross-section of several orders of magnitude (mb, compared to the nb or pb of the previous probes) due to the weak binding of these clusters.
  
 An exotic beam of $^{14}$Be, with the $4n$ threshold at only 5~MeV,
 was sent onto a carbon target at 35~MeV/\edit{nucleon}. Following its breakup, the detection of the $^{10}$Be fragment provided a clean signature of the $4n$ channel.
 For the detection of a potentially liberated tetraneutron cluster, the principle was similar to that used by Chadwick in the discovery of the neutron \cite{Chadwick32}: deduce the mass of the neutral particle from the recoil induced by elastic scattering on charged particles.
 The recoil energy $E_p$ of a proton in the organic scintillator detector is related to the energy per nucleon $E_n$ of the incoming $^A$n cluster, obtained from its time of flight: $(E_p/E_n)\leqslant4A^2/(A+1)^2$.
 Of course, for single neutrons $(E_p/E_n)\leqslant1$, although due to the experimental resolutions events may go up to 1.4.
 Since the dineutron is unbound, and the trineutron should also be due to pairing, the measurement of proton recoils of 1.5--2.5 over the incoming neutron energy could only be attributed to a bound tetraneutron.
 
\begin{figure}[ht] \begin{center}
  \psfig{file=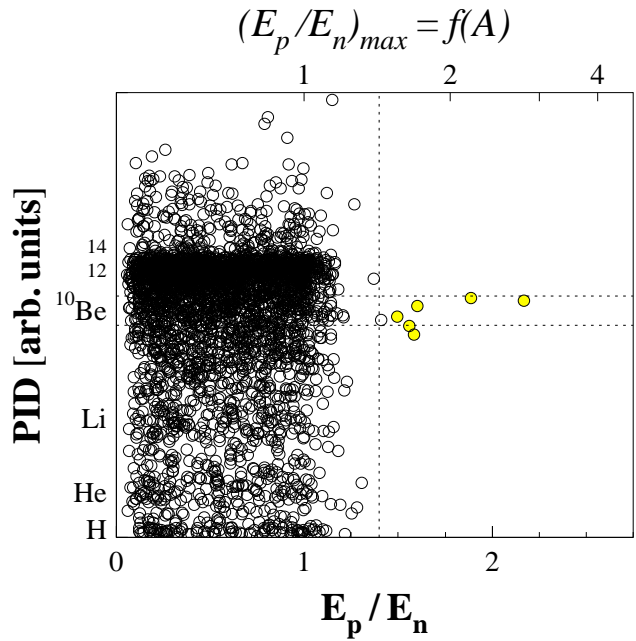,width=8.5cm}
 \end{center} \caption{Scatter plot of the particle identification parameter PID vs the proton recoil in the neutron detector (normalized to the neutron energy) for the reaction $^{14}$Be$\,\rightarrow X+n$. The dotted lines show the region centered on the $^{10}$Be peak and with $E_p/E_n>1.4$, and the 6~events in yellow are the candidates to the formation of a bound tetraneutron. The scale on the upper axis shows the maximum proton recoil as a function of the multineutron mass number (see the text). Adapted from \cite{Marques02_4n_recoil}.} \label{f:GANIL}       
\end{figure}

 The method was also applied to data from $^{11}$Li and $^{15}$B beams, but only in the case of $^{14}$Be some events were observed with characteristics consistent with the production and detection of a tetraneutron. The 6~events are shown in Fig.~\ref{f:GANIL} in yellow, with proton recoils 1.4--2.2 times higher than those expected for individual neutrons, and appear all in coincidence with the detection of a $^{10}$Be fragment (note the absence of candidate events in the most abundant $^{12}$Be channel). 
 Special care was taken to estimate the effects of pileup, i.e.\ the detection of several neutrons in the same detector, and it was found that it could at most account for some 10\% of the observed signal. 
 
 This result triggered several theoretical calculations that could not explain the possible binding of the tetraneutron. Moreover, another work questioned the probe itself \cite{Sherrill_Bertulani}, arguing that a weakly bound tetraneutron would rather undergo breakup. 
 Marqu\'es \etal\ addressed both issues \cite{Marques05_4n_recoil}, finding that the signal observed 
 could be generated also by a low-energy tetraneutron resonance ($E\lesssim2$~MeV) through an enhancement of pileup,
 or by the breakup of a bound tetraneutron in the scintillator followed by the detection of some of the neutrons.
 The decrease in beam intensities at GANIL and the aging of their neutron detector used did not allow the authors to obtain an unambiguous confirmation of this result.
 After some theoretical works in the early 2000s, the field became quiet.

\subsection{The RIKEN result} \label{sec:RIKEN16}

 In 2016, Kisamori \etal\ proposed a new probe at RIKEN: $^4$He$(^8$He$,^8$Be$)^4$n, a DCX reaction using exotic nuclei \cite{Kisamori16_4n_DCX}.
 Sending a very intense $^8$He beam at 186~MeV/\edit{nucleon} onto a liquid $^4$He target, the exit channel was selected through the detection in an spectrometer of the two $\alpha$ particles from the decay in flight of $^8$Be.
 The large $Q$ value of the $(^8$He$,^8$Be$)$ reaction almost compensated the binding energy of the $\alpha$ particle and allowed for the formation of a $4n$ system with small momentum transfer.
 The authors expected in this way to enhance the odds of a weakly interacting tetraneutron in the final state, that would remain in the target area (and would not be directly detected).

\begin{figure}[ht] \begin{center}
  \psfig{file=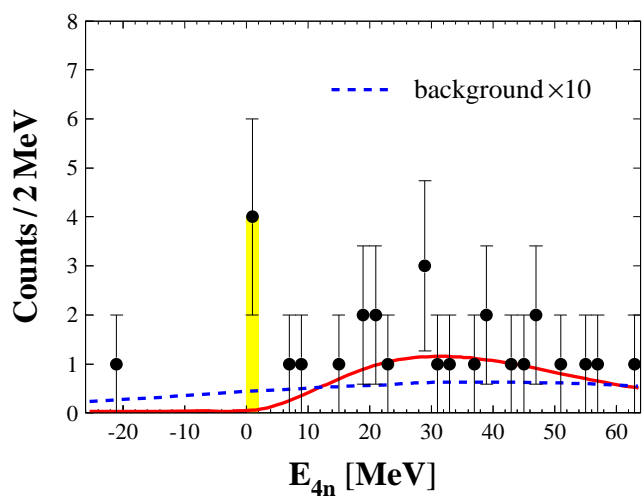,width=8.5cm}
 \end{center} \caption{Missing-mass spectrum of the $^4$He$(^8$He$,^8$Be$)4n$ reaction. The solid (red) curve represents the sum of the direct decay of correlated $2n$ pairs plus the estimated background. The dashed (blue) curve represents only the latter, multiplied by a factor of 10 in order to make it visible.
 The 4~events at threshold are highlighted in yellow.
 Adapted from \cite{Kisamori16_4n_DCX}.} \label{f:Kisamori}
\end{figure}

 The $4n$ missing-mass spectrum (Fig.~\ref{f:Kisamori}) showed 4~events very close to threshold.
 The relative energy and angle between the two $\alpha$ particles were consistent with the formation of the $^8$Be ground state. The background was estimated from the probability of having two $^8$He beam particles from the same bunch breaking up, due to the high beam intensity, and leading to the detection of two independent $\alpha$ particles at small angle that could mimic a $^8$Be decay. It was found to be uniformly spread over the whole range of missing mass (blue curve in Fig.~\ref{f:Kisamori}), with an integrated value of about 2~events.
 The red curve in Fig.~\ref{f:Kisamori} corresponds to a calculation of the direct decay of the $4n$ final state within a wave packet similar to the initial $^4$He, including the interaction between neutrons and between neutron pairs \cite{Kisamori16_4n_DCX}. It includes also the estimated background described above, not visible at the scale of the figure.
 This curve, without the hypothesis of a tetraneutron resonance state, clearly cannot explain the events observed around the $4n$ threshold.

 Note that there was only 1~event in the kinematically forbidden region ($E_{4n}\sim-20$~MeV), a clear indication of the low background.
 The 4~events in the region $0<E_{4n}<2$~MeV were found to be consistent with the formation of a tetraneutron resonance at $E(^4$n$)=0.8\pm1.3$~MeV, with a width $\Gamma<2.6$~MeV, taking into account both the statistic and systematic errors. It should be noted that, because of the missing-mass uncertainty, this result is also consistent with the formation of a bound tetraneutron. The cross-section 
 was estimated \edit{to be} about 4~nb.

\section{What do theories say?} \label{Sec_Th}

 For a complete and detailed review of all the theoretical calculations concerning multineutrons the reader should check Ref.~\cite{EPJA_review}, and will find each of the different works in the references therein. In this section, I will only describe the main categories, cite a few relevant publications, and summarize their conclusions.
 
 The overall consensus of all the calculations is that the known $nn$ interaction cannot bind a $3n$ or $4n$ system. In order to bind them, one needs to multiply $V_{nn}$ by an unrealistically huge factor, incompatible with the basic knowledge of the nucleon-nucleon interactions, and with the description of all light nuclei.
 Therefore, very early, the theory faced the next question: can few neutrons form a resonant state? If so, would it be possible to measure it in a nuclear reaction in the laboratory? However, calculating unbound states is not easy, and the basic principle followed by most of the theories was to bind the unbound system by means of some artifact and then release it progressively and control its evolution in the continuum.

\subsection{Binding (and unbinding) a resonance}

 Even if the principle is the same, its application differed among the different groups, as well as the methods used to solve the many-body problem.
 Of particular importance was the pioneering work of Gl\"ockle \etal, with a series of papers solving exactly the Faddeev equations in momentum space for the $3n$ system \cite{Glockle_3n_PRC18_1978,OG_3n_NPA318_1979,WG_3n_PRC60_1999,HGK_3n_PRC66_2002}.
 They first found that $V_{nn}$ had to be 4.2 times stronger in order to bind the trineutron \cite{Glockle_3n_PRC18_1978}. Then, the addition of higher waves beyond the S-wave lowered this factor, but it was still 3.7 \cite{OG_3n_NPA318_1979}. And of course, in both cases the dineutron became strongly bound, and therefore their artificially barely-bound trineutron would indeed be unbound and decay into $^2$n$+n$.
 
\begin{figure}[ht] \begin{center}
  \psfig{file=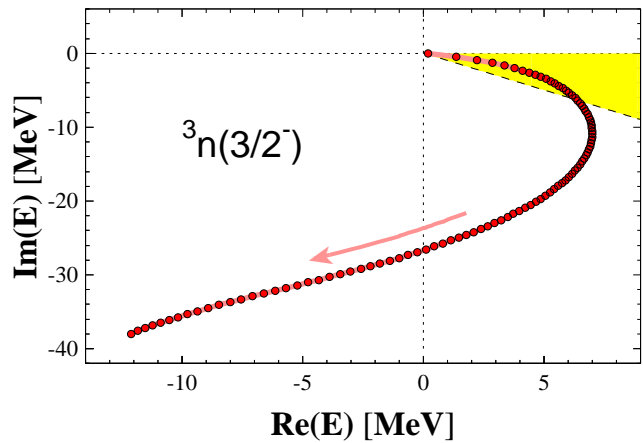,width=8.5cm}
\caption{$S$-matrix pole trajectory of $3n$ for $J^\pi=3/2^-$ in the complex energy plane, from an artificial bound state (with magnified $V_{nn}$ interactions, at the origin) up to the physical case (without scaling, at the end point). Observable resonances are located in the first sector of the second quadrant, highlighted in yellow. Adapted from \cite{HGK_3n_PRC66_2002}.} \label{3n_Gloeckle_32m}       
\end{center}
\end{figure}
 
 They solved this issue in the last works of the series \cite{WG_3n_PRC60_1999,HGK_3n_PRC66_2002}. They did not modify the S-wave, so that the dineutron remained unbound, and then scaled the rest of the waves in order to bind the trineutron. Finally they reduced the scaling factors progressively and followed the trajectory of the state in the complex-energy plane starting from (0,0), as we can see in Fig.~\ref{3n_Gloeckle_32m}.
 As a reference, observable resonances are located in the first sector of the second quadrant, between the dashed line and the real axis, with widths (imaginary part) comparable or smaller than the resonance energy (real part). The $3n$ resonances were all found in the third quadrant, extremely far away from the real axis.
  The overall conclusion was that $3n$ resonances could not exist near the physical region.

 After the renewed interest in the field following the publication of GANIL result, the work of Gl\"ockle was extended by Lazauskas and Carbonell towards more accurate methods to evolve a bound state into the continuum, for the trineutron \cite{LC_3n_PRC71_2005} and tetraneutron \cite{LC_4n_PRC72_2005}.
 Basically, to avoid the artificial binding of subsystems that would lead to nonsense conclusions, they added only a three- or four-body force respectively to the $3n$ and $4n$ systems. This artificial force would not act thus on the subsystems, and could be progressively removed without introducing spurious effects.
 The trajectories they found were similar to the one in Fig.~\ref{3n_Gloeckle_32m}. Therefore, the ensemble of these exact calculations concluded that the trineutron and tetraneutron could not exist either as bound or observable resonant states.

 Following GANIL publication, another type of calculation was carried out by Pieper \cite{Pieper:2003dc}. He applied Green Function Monte-Carlo (GFMC) methods, which had been very successful in reproducing the binding energies of $A=2$--10 nuclei, to study the possible existence of a bound tetraneutron. 
 The conclusion was the same \edit{as} in the previous works: any attempt to force \edit{the formation of} a $^4$n bound state would have ``devastating effects'' in the description of the nuclear chart.
 However, in contrast with those previous works, Pieper surprisingly claimed a possible $^4$n resonance at only about 2~MeV.
 His two conclusions seemed somehow contradictory: the $4n$ Hamiltonian could hardly accommodate at the same time the absolute impossibility of a bound state and the possibility of a near-threshold resonance.

\begin{figure}[ht] \begin{center}
  \psfig{file=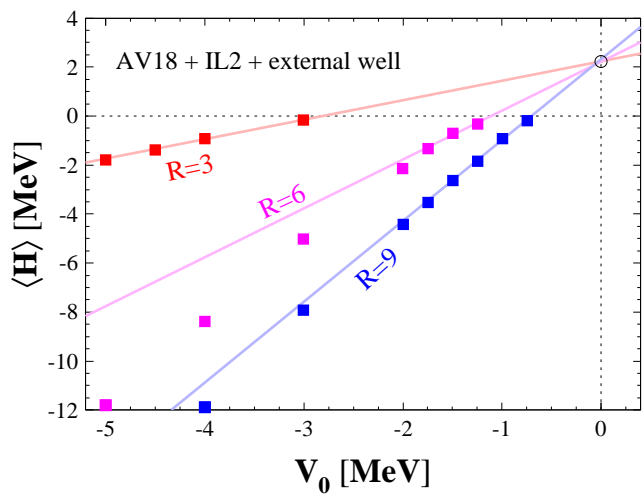,width=8.5cm}
\caption{Energies of $4n$ in a Woods-Saxon trap as a function of the trap strength $V_0$. The results obtained with several values of the trap radius $R$ are used to extrapolate linearly into the continuum, towards the removal of the trap. Adapted from \cite{Pieper:2003dc}.}\label{E4n_WS_Trap}       
\end{center}
\end{figure}
 
  The possible resonant state was determined by computing the energy of $^4$n bound states in a Woods-Saxon trap with fixed diffuseness (Fig.~\ref{E4n_WS_Trap}). Using different values of the trap radius $R$, the trap strength $V_0$ was progressively decreased until the $4n$ became unbound, and the last energy values were extrapolated towards the absence of trap at $V_0=0$. All radii led to similar extrapolated values of $E\sim2$~MeV.
 Pieper's work had a great influence in this topic and it was taken as a definitive proof to justify further experimental, and some theoretical, investigations.
 It is worth noting, however, that his own statement was very cautious: ``since the GFMC calculation with no external well shows no indication of stabilizing at that energy, the resonance, if it exists at all, must be very broad''.

\subsection{Opening and closing (again) the door}

 With his rough approximation, Pieper kept the door to a $4n$ resonance open, a door that had been `exactly' closed by Gl\"ockle and locked by Lazauskas and Carbonell. In that context, the publication of RIKEN result claiming the observation of a resonance finally opened wide that door, and several calculations used similarly rough approximations to extrapolate bound states into observable low-energy resonances \cite{Shirokov:2016ywq,Gandolfi:2016bth,Marek_4n_2017,Li_Michel_3n_PRC100_2019}.
 Besides the approximations used in solving the many-body problem, all of them did bind the `resonances' with either external traps or global scaling factors, which in any event led to systems containing unphysical strongly bound dineutrons, and even trineutrons. Moreover, they all used linear or parabolic extrapolations into the continuum, very far from the complicated paths followed by the poles of the $S$-matrix in the complex-energy plane.
 
  A series of calculations closed again the door, if it had been open at all, to any tri- or tetraneutron state. 
 Right after the publication of RIKEN result, Hiyama \etal\ made a complete study of both systems from the perspective of the almost unknown $T=3/2$ component of the three-body force, that would act on $3n$ and $4n$ \cite{HLCK_4n_PRC93_2016}. Using the same exact methods that Gl\"ockle, Lazauskas and Carbonell had introduced, to solve the many-body system and to evolve bound states into the continuum, they concluded that $3n$ and $4n$ states could not exist either as bound states or observable resonances.
 Moreover, Deltuva and Lazauskas demonstrated rigorously that the resonances predicted by some previous works were in fact artifacts generated by either the trap, the global scaling or the extrapolation methods used \cite{Comment_PRL123_2019,Arnas_Rimas_4n_2n_PRC100_2019}. Very recently, Ishikawa clearly illustrated those artifacts in a $3n$ system inside a trap \cite{Ishikawa_3n_PRC102_2020}.

 If we agree that bound or resonant states cannot exist in light multineutrons, at least for $A=3,4$, how can we explain the signals observed at GANIL and RIKEN? Higgins \etal\ have recently studied the $3n$ and $4n$ systems within the adiabatic hyper-spherical framework \cite{HGKV_PRL125_2020}. The corresponding adiabatic potential energy curve was analyzed and found to be repulsive, and they concluded that there is no sign of a low-energy resonance for any of these systems. However, they observed in both of them some low-energy enhancement of the so-called Wigner-Smith ``time delay'' \cite{HGKV_PRL125_2020}, that could provide a hint to understand the GANIL and RIKEN near-threshold enhancements.
 Interestingly, Deltuva had found that even in the absence of any $4n$ resonance, some $4\to4$ transition operators exhibit a low-energy enhancement \cite{AD_4n_PLB782_2018}. He conjectured that they could also manifest in other reactions with the $4n$ subsystem in the final state, like the ($^{14}$Be,$^{10}$Be) reaction at GANIL or the $^4$He($^8$He$,^8$Be) one at RIKEN.
 
 The door may be still open, but not for what we expected?

\section{Present and future projects} \label{s:future}

\begin{figure}[ht] \begin{center}
  \psfig{file=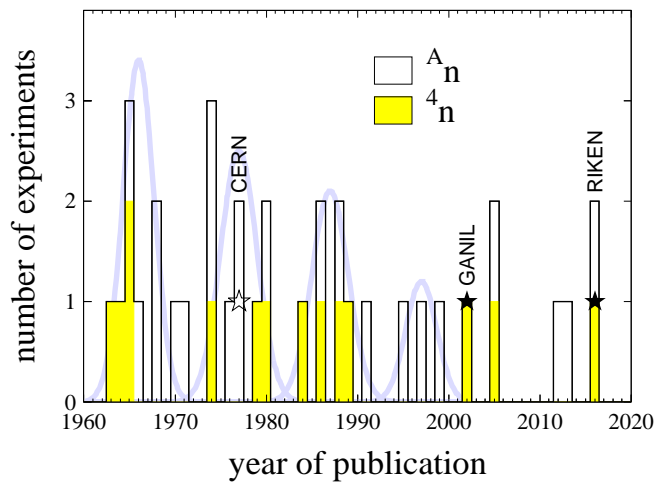,width=8.5cm}
 \end{center} \caption{The solid histogram represents the number of experiments searching for multineutrons (\edit{36 total}, mainly for tri- and tetraneutrons) as a function of the year of publication of the results. In yellow are those searching specifically for the tetraneutron.
 The stars represent the three positive signals reported, the empty one that was refuted \cite{Detraz77_68n_pW} and the two solid ones that have not been contested yet \cite{Marques02_4n_recoil,Kisamori16_4n_DCX}.
 \edit{The pale Gaussians guide the eye through the recurring pattern.}} \label{f:history}       
\end{figure}

 In this sixty-year period, 36 experiments (with results published in internationally accessible journals) have searched for evidence of multineutron existence. Their chronology (Fig.~\ref{f:history}) exhibits several trends.
 Even if the techniques have been diverse and their sensitivity has increased with time, we can see a recurring pattern of `bunches' during the first forty years, with experiments accumulating from mid to end of each decade. Some experiments at mid-decade triggered others, and then the overall negative results lead to a temporary stop in the program, until someone else restarted it a few years later.
 Towards the end of the century the number of experiments in each bunch decreased, showing signs of exhaustion due to the lack of positive signals.

 The number of experiments in the present century has been much lower, although two positive signals of a bound or low-lying resonant tetraneutron were obtained. 
 As said in the introduction, those signals have renewed the interest in the field, both experimentally and theoretically. Therefore, for the next extension of Fig.~\ref{f:history} we expect an upcoming significant `bunch' of results.

\subsection{Detecting the neutrons at last}

 As mentioned in Sec.~\ref{s:production}, paradoxically, none of the many experiments already discussed has tried to detect the neutrons, although we note that the GANIL result was interpreted as the direct detection of a cluster of neutrons.
 With the recent improvements in beam intensities and neutron detection efficiencies in the world-leading facilities, trying to detect the neutrons from the decay of a resonance or the breakup of a weakly bound state seems a logical next step, which in addition would give access to the eventual correlations in the system (Fig.~\ref{f:categories}c).

 For the sake of completeness, I will mention that in 2016 Bystritsky \etal\ claimed the first direct detection of multineutrons with $A=6,8$ \cite{Bystritsky16_6n_natU}. They used an array of 
 $^3$He counters, with a total estimated efficiency of $\varepsilon_{1n}\sim20\%$, surrounding a $^{238}$U sample, and searched for the neutrons that would stem from the multineutron cluster decay of uranium.
 Although they based their claim in the detection of few events with 5~neutron hits, they also admitted that the confirmation would require an increase of the statistics, and a reduction of the cosmic and natural backgrounds, by at least one order of magnitude \cite{Bystritsky16_6n_natU}.
 
 The Radioactive Isotope Beam Factory (RIBF) of the RIKEN Nishina Center provides nowadays the highest intensities of light neutron-rich beams, together with high neutron-detection efficiencies.
 Since the 2016 result, two experiments have been undertaken at RIKEN aiming at the detection, for the first time, of all the neutrons emitted in a multineutron ($x>2$) decay.
 They benefited both from the combined capabilities of the NEBULA \cite{NEBULA} and NeuLAND demonstrator \cite{NeuLAND} neutron detector arrays. With an average neutron efficiency of $\varepsilon_{1n}\sim45\%$, the granularity of the arrays allowed for an unprecedented four-neutron efficiency of $\varepsilon_{4n}\sim1$--2\% (see Fig.~\ref{f:efficiency}).

 The first experiment searched for the ground state of $^{28}$O, one of the grails of nuclear structure physics, using the reaction H$(^{29}$F$,^{24}$O$)4n$ \cite{NP_Kondo}. The moderate intensity of the very exotic $^{29}$F beam lead to some 100 complete $4n$ events. Although the analysis of the results is still in progress, it seems that the decay of the ground state of $^{28}$O would proceed through the narrow $^{26}$O ground state, i.e.\ it would be a sequential $2n$-$2n$ decay. As such, no information should be derived concerning multineutrons.
 
 The second experiment used a similar setup and technique, the proton removal from a beam and the detection of a fragment plus four neutrons, with the reaction H$(^8$He$,^3$H$)4n$ \cite{NP_Marques}. The goal was twofold: the precise location of the $^7$H ground state, which as we saw in Fig.~\ref{f:masses} could be the key to the multineutron puzzle; and the search for its potential tetraneutron decay. With respect to the previous experiment, the much higher intensity of the $^8$He beam will lead to several orders of magnitude increase in statistics, providing tens of thousands of $4n$ events, while the absence of low-lying $^{4,5,6}$H resonances should allow for the observation of a direct $4n$ decay, and the correlations within.

\subsection{The end may be near?}

 Moreover, besides those two experiments aiming at the detection of four neutrons already carried out \cite{NP_Kondo,NP_Marques}, complementary missing-mass experiments are also being programmed at RIKEN, without neutron detection but with increased sensitivity.
 The tetraneutron has already been revisited using the same reaction as in 2016, $^4$He$(^8$He$,^8$Be$)^4$n, with several improvements in the experimental conditions \cite{NP_Shimoura}. The tetraneutron has also been probed in the knockout of an $\alpha$ particle off $^8$He at backward angles in quasi-free conditions, H$(^8$He$,p\alpha)^4$n \cite{NP_Paschalis}, avoiding the FSI of the $4n$ with the other particles in the reaction.
 
 Among those missing-mass experiments, some are already searching for the next heavier system, the hexaneutron. One has already been undertaken, knocking out two $\alpha$ particles from $^{14}$Be in the reactions H$(^{12,14}$Be$,p\alpha\alpha)^{4,6}$n \cite{NP_Beaumel}, and a second one is planned in a near future, knocking out an $\alpha$ particle and a proton from $^{11}$Li in the reaction H$(^{11}$Li$,pp\alpha)^6$n \cite{NP_Nakamura}. Depending on their results, future experiments could be planned in order to study neutron correlations in the decay of hexaneutron states.

 Therefore, taking into account the increasing accuracy and sensitivity of these new experiments, the next few years will possibly see the end of this quest whatever the outcome, at least with respect to $^{3,4}$n.
 However, the hexaneutron seems to represent a mass frontier difficult to cross in the laboratory. Theoretical calculations could help us go beyond $A=6$ in order to understand possible binding energy trends in these systems.

\section{Summary and perspectives}

 The multineutron quest is full of controversies and debates, but it is definitely a fascinating story. Already since the distant 1960s, experimentalists have been playing with pion beams, approaching nuclear reactors, guiding exotic nuclei with powerful accelerators... trying to create a grail against all odds: neutron matter in the laboratory.
 In parallel, theoreticians have been improving the first three-body calculations, and are presently solving exactly the five-body problem.
 Sixty years later, the quest is still fascinating but still open. If from the theoretical point of view the question of a narrow $3n$ and $4n$ resonance seems closed, there is an urgent need for a definite experimental conclusion concerning the GANIL and RIKEN events.

 Nuclear theory, and in particular the description of light nuclei from the underlying forces between nucleons, has seen an enormous progress working hand in hand with experiments. The main theoretical models have been born from the known properties of stable nuclei, and then they have been refined and updated with the increasing knowledge about more and more exotic isotopes.
 In the field of neutral nuclei, however, the theory advances almost blind.
 The only two experimental signals \cite{Marques02_4n_recoil,Kisamori16_4n_DCX} are still weak and do not provide any firm and precise observable that could benchmark the different methods and techniques.

 In this respect, it should be a priority to confront the controversies surrounding the tri- and tetraneutron calculations with the help of high-statistics, unambiguous experimental results. New techniques or experiments that would provide still another weak and/or ambiguous signal do not seem the best way to unlock progress in the field.
 The new experiments already (or soon to be) undertaken at RIKEN may bring the very much needed reliable reference points for the theoretical conclusions to be adapted.
 If the positive signals were refuted, then the calculations predicting resonances should be reevaluated. If they were confirmed though, then the detailed characteristics of the signals (energies, widths or others) should be described. 

 Once the $3n$ and $4n$ systems will be clarified, both from theory and experiment, it will be easier to move on to the heavier systems, in particular $6n$ and $8n$.
 Even if the lightest multineutron states were not observable, a pertinent question remains: when increasing the mass number, can at some point a multineutron manifest as a bound or resonant state? Even for the most reluctant theories, and of course for experiments, this is still an open question.

  The known as ``helium anomaly'', i.e.\ the fact that $^8$He is more bound than $^6$He, has long been a clue towards the search for light multineutrons, taken as a hint for additional binding due to the increasing number of neutrons on top of $^4$He.
 Less known but maybe more spectacular would be the analog ``hydrogen anomaly'', provided it is established (Fig.~\ref{f:masses}).
 The $^{4,5,6}$H isotopes are unbound by several MeV and exhibit broad resonances, but $^5$H ground state is narrower than the other two, and $^7$H is reported to lie almost at the $4n$ threshold \cite{H7_GANIL_2007}, although it has in common with the tetraneutron that the experimental signals to date are weak, ambiguous and sometimes contradictory.

 With the recent progress in the exact calculation of the five-body problem, $^7$H is now accessible almost {\it ab initio} (treating the triton as a particle) for the theory and, as we have seen at the end of Sec.~\ref{s:future}, a related experiment that should provide very high statistics and resolution has been already undertaken.
 This super-heavy isotope of hydrogen could in this way be the key for the next steps in the field, both for the role of the tetraneutron at the $t+4n$ threshold but even for the hexaneutron at the $p+6n$ threshold.
 Going beyond $A=6$ represents still a too important obstacle for experiments, and for the moment should be left for the theory, once the lighter multineutrons have helped to provide a solid base for the models.

 In any event, this domain will remain fascinating since, paraphrasing
the authors of Ref.~\cite{Marek_4n_2017}, it ``would deeply impact our
understanding of nuclear matter and requires a critical investigation''.


\begin{thebibliography}{99}

\bibitem{Machleidt_2001} R.~Machleidt, Nucl.\ Phys.\ A {\bf689}, 11c (2001); R.~Machleidt and I.~Slaus, J.\ Phys.\ G {\bf27}, R69 (2001).
\bibitem{Epelbaum_2020} E.~Epelbaum \etal, Eur.\ Phys.\ J.\ A {\bf56}, 92 (2020).
\bibitem{Sekiguchi_2002} K.~Sekiguchi et al., Phys.\ Rev.\ C {\bf65}, 034003 (2002).
\bibitem{Sekiguchi_2004} K.~Sekiguchi et al., Phys.\ Rev.\ C {\bf70}, 014001 (2004).
\bibitem{Povh_1999} B.~Povh \etal, {\em Particles and Nuclei: an Introduction to the Physical Concepts}, Springer (1999).
\bibitem{Haloes} I.~Tanihata \etal, Prog.\ Part.\ Nucl.\ Phys.\ {\bf68}, 215 (2013).
\bibitem{Molecules} M.~Freer \etal, Rev.\ Mod.\ Phys.\ {\bf90}, 035004 (2018).
\bibitem{Hornyak_1975} W.F.~Hornyak, {\em Nuclear Structure}, Academic Press (1975).
\bibitem{AME2016}	M.~Wang \etal, Chin.\ Phys.\ C {\bf41}, 030003 (2017).
\bibitem{Guardiola_PRL84_2001}	R.~Guardiola and J.~Navarro, Phys.\ Rev.\ Lett.\ {\bf84}, 1144 (2000).
\bibitem{He3} R.A.~Aziz and M.J.~Slaman, J.\ Chem.\ Phys.\ {\bf94}, 8047 (1991).
\bibitem{Vnn} V.G.J.~Stoks \etal, Phys.\ Rev.\ C {\bf48}, 792 (1993).

\bibitem{4n_stars_USAL_EJPA155_2019} O.~Ivanytskyi \etal, Eur.\ Phys.\ J.\ A {\bf55}, 184 (2019).
\bibitem{Ogloblin89} A.A.~Ogloblin and Y.E.~Penionzhkevich, in {\em Nuclei Far From Stability, Treatise on Heavy-Ion Science}, edited by D.A.~Bromley (Plenum, New York, 1989), Vol.~8, p.~261, and references therein.
\bibitem{Marques02_4n_recoil} F.M.~Marqu\'es \etal, Phys.\ Rev.\ C {\bf
65}, 044006 (2002).
\bibitem{Kisamori16_4n_DCX}  K.~Kisamori \etal, Phys.\ Rev.\ Lett.\ {\bf
116}, 052501 (2016).
\bibitem{Chadwick32}	J.~Chadwick, Nature {\bf129}, 312 (1932).
\bibitem{cross-talk}	F.M.~Marqu\'es \etal, Nucl.\ Instrum.\ Methods Phys.\ Res.\ {\bf A450}, 109 (2000).

\bibitem{Schiffer63_4n_nU}	J.P.~Schiffer and R.~Vandenbosch, Phys.\ Lett.\ {\bf5}, 292 (1963).
\bibitem{Cierjacks65_4n_d238U}	S.~Cierjacks \etal, Phys.\ Rev.\ {\bf137}, B345 (1965).
\bibitem{Detraz77_68n_pW}	C.~D\'etraz, Phys.\ Lett.\ {\bf66B}, 333 (1977).
\bibitem{Turkevich77_6n_pU}	A.~Turkevich \etal, Phys.\ Rev.\ Lett.\ {\bf38}, 1129 (1977).
\bibitem{deBoer80_46n_3He130Te}	F.W.N.~de Boer \etal, Nucl.\ Phys.\ {\bf A350}, 149 (1980).
\bibitem{Novatsky12_6n_a238U}	B.G.~Novatsky \etal, JETP Lett.\ {\bf96}, No.~5, 280 (2012).
\bibitem{Novatsky13_6n_a238U}	B.G.~Novatsky \etal, JETP Lett.\ {\bf98}, No.~11, 656 (2013).
\bibitem{Gilly65_4n_pi4He}	L.~Gilly \etal, Phys.\ Lett.\ {\bf19}, 335
(1965).
\bibitem{Sperinde70_3n_pi3He}	J.~Sperinde \etal, Phys.\ Lett.\ {\bf32B}, 185 (1970).
\bibitem{Sperinde74_3n_pi3He}	J.~Sperinde \etal, Nucl.\ Phys.\ {\bf B78}, 345 (1974).
\bibitem{Jibuti85_34n_pi34He}	R.I.~Jibuti and R.Ya.~Kezerashvili, Nucl.\ Phys.\ {\bf A437}, 687 (1985).
\bibitem{Ungar84_4n_pi4He}	J.E.~Ungar \etal, Phys.\ Lett.\ {\bf144B}, 333 (1984).
\bibitem{Stetz86_34n_pi34He}	A.~Stetz \etal, Nucl.\ Phys.\ {\bf A457}, 669 (1986).
\bibitem{Gorringe89_4n_pi4He}	T.P.~Gorringe \etal, Phys.\ Rev.\ C {\bf40}, 2390 (1989).
\bibitem{Yuly97_3n_pi3He}	M.~Yuly \etal, Phys.\ Rev.\ C {\bf55}, 1848 (1997).
\bibitem{Grater99_3n_pi3He}	J.~Gr\"ater \etal, Eur.\ Phys.\ J.\ B {\bf4}, 5 (1999).
\bibitem{Bistirlich76_3n_pi3H}	J.A.~Bistirlich \etal, Phys.\ Rev.\ Lett.\ {\bf36}, 942 (1976).
\bibitem{Miller80_3n_pi3H}	J.P.~Miller \etal, Nucl.\ Phys.\ {\bf A343}, 341 (1980).
\bibitem{Chultem79_4n_pi208Pb}	D.~Chultem \etal, Nucl.\ Phys.\ {\bf A316}, 290 (1979).
\bibitem{Gornov91_3n_pi9Be}	M.G.~Gornov \etal, Nucl.\ Phys.\ {\bf A531}, 613 (1991).
\bibitem{Ajdacic65_3n_n3H}	V.~Ajda\v{c}i\'c \etal, Phys.\ Rev.\ Lett.\ {\bf14}, 444 (1965).
\bibitem{Thornton66_3n_n3H}	S.T.~Thornton \etal, Phys.\ Rev.\ Lett.\ {\bf17}, 701 (1966).
\bibitem{Ohlsen68_3n_t3H}	G.G.~Ohlsen \etal, Phys.\ Rev.\ {\bf176}, 1163 (1968).
\bibitem{Cerny74_43n_transfer}	J.~Cerny \etal, Phys.\ Lett.\ {\bf53B}, 247 (1974).
\bibitem{Belozyorov88_43n_transfer}	A.V.~Belozyorov \etal, Nucl.\ Phys.\ {\bf A477}, 131 (1988).
\bibitem{Bohlen95_3n_transfer}	H.G.~Bohlen \etal, Nucl.\ Phys.\ {\bf A583}, 775 (1995).
\bibitem{Aleksandrov05_43n_transfer}	D.V.~Aleksandrov \etal, JETP Lett.\ {\bf81}, No.~2, 43 (2005).

\bibitem{InFlight} T.~Nakamura \etal, Prog.\ Part.\ Nucl.\ Phys.\ {\bf97}, 53 (2017).
\bibitem{Marques_19C}	F.M.~Marqu\'es \etal, Phys.\ Lett.\ B {\bf381}, 407 (1996).

\bibitem{Sherrill_Bertulani}	B.M.~Sherrill and C.A.~Bertulani, Phys.\ Rev.\ C {\bf69}, 027601 (2004).
\bibitem{Marques05_4n_recoil}	F.M.~Marqu\'es \etal, arXiv nucl-ex/0504009.

\bibitem{EPJA_review} F.M.~Marqu\'es and J.~Carbonell, Eur.\ Phys.\ J. A {\bf 57}, 105 (2021).

\bibitem{Glockle_3n_PRC18_1978}	W.~Gl\"ockle, Phys.\ Rev.\ C {\bf18}, 564 (1978).
\bibitem{OG_3n_NPA318_1979}	R.~Offermann and W.~Gl\"ockle, Nucl.\ Phys.\ A {\bf318}, 138 (1979).
\bibitem{WG_3n_PRC60_1999}	H.~Wita\l{}a and W.~Gl\"ockle, Phys.\ Rev.\ C {\bf60}, 024002 (1999).
\bibitem{HGK_3n_PRC66_2002}	A.~Hemmdan \etal, Phys.\ Rev.\ C {\bf66}, 054001 (2002).
\bibitem{LC_3n_PRC71_2005}	R.~Lazauskas and J.~Carbonell, Phys.\ Rev.\ C {\bf71}, 044004 (2005).
\bibitem{LC_4n_PRC72_2005}	R.~Lazauskas and J.~Carbonell, Phys.\ Rev.\ C {\bf72}, 034003 (2005).
\bibitem{Pieper:2003dc}	S.C.~Pieper, Phys.\ Rev.\ Lett.\ {\bf90},  252501 (2003).

\bibitem{Shirokov:2016ywq}	A.M.~Shirokov \etal, Phys.\ Rev.\ Lett.\ {\bf117}, 182502 (2016).
\bibitem{Gandolfi:2016bth}	S.~Gandolfi \etal, Phys.\ Rev.\ Lett.\ {\bf118}, 232501 (2017).
\bibitem{Marek_4n_2017}	K.~Fossez \etal, Phys.\ Rev.\ Lett.\ {\bf119}, 032501 (2017). 
\bibitem{Li_Michel_3n_PRC100_2019}	J.G.~Li \etal, Phys.\ Rev.\ C {\bf100}, 054313 (2019).

\bibitem{HLCK_4n_PRC93_2016}E.~Hiyama \etal, Phys.\ Rev.\ C {\bf93}, 044004 (2016).
\bibitem{Comment_PRL123_2019}	A.~Deltuva and R.~Lazauskas, Phys.\ Rev.\ Lett.\ {\bf123}, 069201 (2019).
\bibitem{Arnas_Rimas_4n_2n_PRC100_2019}	A.~Deltuva and R.~Lazauskas, Phys.\ Rev.\ C {\bf100}, 044002 (2019).
\bibitem{Ishikawa_3n_PRC102_2020}	S.~Ishikawa, Phys.\ Rev.\ C {\bf102}, 034002 (2020).

\bibitem{HGKV_PRL125_2020}	M.D.~Higgins \etal, Phys.\ Rev.\ Lett.\ {\bf125}, 052501 (2020).
\bibitem{AD_4n_PLB782_2018}	A.~Deltuva, Phys.\ Lett.\ B {\bf782}, 238 (2018).

\bibitem{Bystritsky16_6n_natU}	V.M.~Bystritsky \etal, Nucl.\ Instrum.\ Methods Phys.\ Res.\ {\bf A834}, 164 (2016).
\bibitem{NEBULA}	T.~Nakamura and Y.~Kondo, Nucl.\ Instrum.\ Meth. Phys.\ Res. B {\bf376}, 1 (2015).
\bibitem{NeuLAND}	Technical report of NeuLAND, https://edms.cern.ch/ui/file/1865739/2/TDR\_R3B\_NeuLAND\_public.pdf
\bibitem{NP_Kondo}	Y.~Kondo \etal, RIBF Proposal NP1312-SAMURAI21.
\bibitem{NP_Marques}F.M.~Marqu\'es \etal, RIBF Proposal NP1512-SAMURAI34.
\bibitem{NP_Shimoura}	S.~Shimoura \etal, RIBF Proposal NP1512-SHARAQ10.
\bibitem{NP_Paschalis}	S.~Paschalis \etal, RIBF Proposal NP1406-SAMURAI19.
\bibitem{NP_Beaumel}	D.~Beaumel \etal, RIBF Proposal NP1206-SAMURAI12.
\bibitem{NP_Nakamura}	T.~Nakamura \etal, RIBF Proposal NP1812-SAMURAI47.

\bibitem{H7_GANIL_2007}	M.~Caama\~{n}o \etal, Phys.\ Rev.\ Lett.\ {\bf99}, 062502 (2007).

\end{thebibliography}
%

\end{document}